\documentclass[12pt]{amsart}

\usepackage{amsmath,amssymb,psfrag,epsfig,boxedminipage,helvet,theorem,endnotes,version,enumerate,environ,rotating}
\usepackage{ stmaryrd,enumitem }
\newcommand{\remove}[1]{}
\usepackage{graphicx, amsfonts}

\usepackage[myheadings]{fullpage}

\newcommand{\dsum}{\displaystyle \sum}
\newcommand{\Z}{\mathbb{Z}}

\begin{document}


\title{Secure Montgomery Multiplication and Repeated Squares for Modular Exponentiation}
\author{Justin Bloom}
\address{Oregon State University}
\email{bloomju@oregonstate.edu}
\author{Lalita Devadas}
\address{Columbia University}
\email{lalita.devadas@columbia.edu}

\dedicatory
{\textsc{Advisor: Dr. Mike Rosulek\\Oregon State University}}

\thanks{This work was done during the Summer 2019 REU program in Mathematics and Theoretical Computer Science at Oregon State University supported by NSF grant DMS-1757995}
\date{August 16, 2019}

\maketitle

\markboth{Bloom, Devadas, Rosulek}{Secure Montgomery Multiplication and Repeated Squares for Modular Exponentiation}


\section*{Abstract}

The BMR16 circuit garbling scheme \cite{BMR16} introduces gadgets that allow for ciphertext-free modular addition, while the multiplication of private
inputs modulo a prime $p$ can be done with $2(p-1)$ ciphertexts as described in \cite{MPS}.
By using a residue number system (RNS), we can construct a circuit to handle the squaring and multiplication of inputs modulo a large $N$ via the
methods described in \cite{recursiveRes}.
We expand on the existing techniques for arithmetic modulo $p$ to develop methods to handle arithmetic in a positional, base-$p$ number system.
We evaluate the ciphertext cost of both of these methods and compare their performance for squaring in various large moduli.
%
%

\section{Introduction} 
\subsection{Secure Multiparty Computation}
Multiparty computation (MPC) is a process in which two parties, Alice and Bob, have private information $x_1, x_2 \in X$, and
an operation using each party's information $F: X^2 \to Y$ is computed. 
If $y = F(x_1, x_2)$ can somehow be computed and made public so that Alice can not find the value of $x_2$ 
and Bob can not find the value of $x_1$, then this process is called secure MPC. 

Circuit garbling is a technique that provides a means for how a secure MPC scheme can be constructed.
Alice and Bob devise a circuit abstraction of $F$, breaking down the sub-processes needed in the computation of $F$ into gates with input and output wires. 
A gate with $k$ input wires from a space $X$ and $n$ output wires for a space $Y$, and functionality $f : X^k \to Y^n$,
can be naively implemented with $n$ lookup tables (one per output wire). Each lookup table has $|X|^k$ entries
for the possible values held by the input wires. Alice and Bob can flip a coin to decide who shall act as a ``garbler'' and who shall act as an ``evaluator''.

Without trying to be clever, the garbler makes public a set of $n|X|^k$ ciphertexts that an evaluator could decrypt to recover each output of $f$. 
The garbler generates these ciphertexts by encrypting the outputs of $f$ using a key that is an encoding of the appropriate inputs, 
where the code being used satisfies the following:
\begin{enumerate}
	\item The evaluator knows which ciphertext to decrypt given $k$ input wires that hold an encoding of the inputs.
	\item The evaluator is only able to find an authentic plaintext from decrypting the correct ciphertext using
	the key derived from the input wires. 
\end{enumerate} 

We make no assumptions on the physical capabilities of either the evaluator or the garbler, and instead assume
that the bottlenecking factor in the entire garbling protocol is the communication of all necessary ciphertexts from the garbler to the evaluator.
Hence we gauge the efficiency of a garbling scheme by the number of necessary ciphertexts, called the \textit{ciphertext cost}. 

\subsection{Our Results}
By building off of arithmetic garbling techniques by Ball, Malkin and Rosulek \cite{BMR16}, we give two approaches for garbling circuits that
perform modular exponentiation. For the repeated squares fast exponentiation algorithm, we need an efficient means for squaring, and multiplying modulo $N$ and so we explore our options for a unary squaring gate, and a boolean multiplication gate. 
\subsubsection{The RNS Approach}
The first approach is using an RNS with relatively prime moduli $p_i$ for $1\le i \le k$ (called the \textit{native moduli}), where we can use small unary gates that
square an input in a prime modulus to find the square of an input given in RNS representation. Then, using algorithms from \cite{recursiveRes}, given an
RNS representation of some residues $x, y < \prod p_i$ we can
find the RNS representation for $x^2 \mod N$ or $xy \mod N$ for known conditions on $N$, without the costly operation of dividing by $N$. 

Since $N$ is assumed to be large, not only is it computationally expensive to divide by $N$ in our native moduli,
but it would would be infeasible to garble the entirety of a single gate for arithmetic in $\Z_N$, and so we refer to $N$ as the \textit{non-native modulus}. 

Therefore, we can garble a circuit consisting of many arithmetic gates connected to the initital output from squaring in each native modulus, to 
perform the appropriate computation from the algorithms described in \cite{recursiveRes}.
\subsubsection{The Base-$p$ Approach}
We compare the ciphertext cost of the RNS ciruit with our second approach, which uses a positional, base-$p$ number system for some prime $p$.
We first expand the ideas of free modular addition and multiplication by a public constant from \cite{BMR16} to find the cost of base-$p$ addition and multiplication
with single-digit input wires. 

For addition, we find the ciphertext cost of garbling a gate with $m$ input wires for single-digit integers and two output wires representing the
two digits of the output sum. The least significant position is exactly the output of the inputs' sum modulo p,
while the most significant position is computed using the technique in section \ref{MulBasePCosts}. 
Since the output must be two digits (where 0 is possible for either position), we assume $m \le p+1$. 
We use this construction to generalize the cost of garbling a circuit which computes the sum of integers with many digits represented in base-$p$. 

For multiplication, we develop a `base-case' for multiplying two single digit numbers (bounded by $p$) in base-$p$ with two output wires
for the digits of the output (which must be bounded by $p^2$), and we simplify our technique for a similar single-digit base-case for squaring. 
Then using the classical divide-and-conquer Karatsuba algorithm, we find the ciphertext cost of multiplying and squaring for any digit base-$p$ representations. 

\section{Background and Notation}
We will take a look at garbled circuit constructions, or ``gadgets'' to build off of, and some techniques in arithmetic that
motivate our results. We denote the ring of integers modulo $m$ by $\Z_m$, and we denote the direct product of rings $R$ and $R'$ by $R \times R'$. 
For an integer $x$, we denote the unique residue $x \mod m$ in $[0, m)$ by $[x]_m$ unless stated otherwise. 

\subsection{Current Garbling Schemes}
The details of how the ciphertext cost of a circuit can be improved are usually in how the possible values of a gate are encoded.
Ball, Malkin and Rosulek \cite{BMR16} developed an encoding technique that allows for addition in a fixed modulus, without any
ciphertexts needed. The evaluator only needs the encodings of two vectors over $\Z_m$ to find the encoding of their sum, and so an adder-gate
can be garbled for free. Using a similar technique, a unary gate that multiplies an input in $\Z_m$ by a public constant can be garbled for free if the constant
is relatively prime to $m$. 

For computations involving modular arithmetic in multiple moduli, a ``mixed-modulus simple circuit'' can be constructed by using an additional
kind of unary gate, that \cite{BMR16} calls ``projection''. A gate representing a projection from $\Z_m$ to $\Z_n$ can be garbled with $m-1$ ciphertexts. 

Finally, for a prime modulus $p$, Malkin, Pastro and Shelat \cite{MPS} 
describe a technique for multiplying two private inputs in $\Z_p$ which costs $2(p-1)$

\subsection{Classical Montgomery Arithmetic}\footnote{Here, we refer to base-$p$ Montgomery algorithms as ``Classical Montgomery Arithmetic'' to distinguish them
from the RNS algorithms, which are also named after Peter Montgomery.}
Despite the improvements made for efficient multiplication, division generally remains costly, and so
repeated division by a large, non-native modulus $N$ is not an efficient way to produce residues modulo $N$. Instead, 
we use classical Montgomery Arithmetic on numbers represented in base-$p$ to perform addition and multiplication modulo $N$ without costly division \cite{smart}. 
To do this, we take advantage of trivial multiplication and division by powers of $p$ for numbers represented in base-$p$.  

\subsubsection{Montgomery Representation}
We take $M$ to be the native \textit{Montgomery Modulus} which is chosen to be $p^k$ for some $k$. 
To do arithmetic on inputs $X$ and $Y$ represented in base-$p$, we first compute the base-$p$ representation for 
the residues of $x := XM \mod N$ and $y := YM \mod N$, called
the \textit{Montgomery Representations} of $X$ and $Y$ respectively. These representations are pre-computed with a division by $N$ each, but
given inputs already in Montgomery representation, it is easy to find the montgomery representation of operations using inputs, without division by $N$.
For addition, we can simply take the sum $x+y = XM + YM \equiv (X+Y)M \mod N$, to find the montgomery represenation of $(X+Y)$. 
\subsubsection{Multiplication}
Multiplying two inputs in Montgomery representation yields an extra factor of M, and so instead we use a technique to compute $xy/M = (XM)(YM)/M = XYM$. 

First we precompute $q := N^{-1} \mod M$ by the Euclidean algorithm, and we take $w := xy$. Then we compute the residue $u :=  -wq \mod M$ by taking the last $k$ digits
of $nM - wq$ for a sufficiently large $n$. Then $(w + uN) \equiv w-wqN \equiv w(1-qN) \equiv 0 \mod M$, 
and therefore the least significant $k$ digits of $(w+uN)$ are 0 in base-$p$.
Furthermore, $(w+uN) \equiv w \mod N$ and therefore truncating $k$ zeros from (w+uN) is equivalent to taking $(w+uN)M^{-1} \equiv wM^{-1} \mod N$.

It is important to note that $(w+uN)/M$ does not necessarily produce a residue bounded by $N$, because we can only guarentee that
$w + uN < N^2 + MN \le 2NM$, and so $(w+uN)/M < 2N$. 

\subsection{Residue Number Systems}
For relatively prime integers $p_1, p_2, ..., p_k$, let $M = \prod_{i=1}^k p_i$. Then the Chinese Remainder Theorem induces an isomorphism 
$\varphi: \Z_M \to \Z_{p_1} \times \Z_{p_2} \times ... \times \Z_{p_k}$, where $\varphi: [x]_M \mapsto ([x]_{p_1}, [x]_{p_2}, ..., [x]_{p_k})$. A \textit{Residue Number System} (RNS) is a way to represent $x$, a residue modulo $M$,
by a vector of residues given by $\varphi(x)$, so that arithmetic in $\Z_M$ can be delegated to $k$ operations in each $p_i$, which can be done in parallel.
We use $b = (p_1, p_2, ... , p_k)$ to denote an RNS with moduli $p_i$. We call $M = \prod_{i=1}^k p_i$ the \textit{dynamical range} of $b$,
since essentially any operation over the integers which does not exceed the dynamical range $M$ can be done in $b$ so that the output can be fully recovered. We use $\llbracket x\rrbracket_b$ to denote the representation of $x$ in the RNS $b$. 

Hollmann, Rietman, Hoogh, Tolhuizen, and Gorissen \cite{recursiveRes} developed a recursive RNS algorithm to perform arithmetic in a non-native modulus.
We provide a ciphertext cost analysis for a garbled circuit implementation of \cite{recursiveRes} in section \ref{RNSCosts}. 

\subsection{Positional Number Systems}
An integer $x$ satisfying $0 \le x \le p^k-1$ can be represented in a positional, base-$p$ number system with $k$ positions, or digits. 
We choose a prime number $p$ as our base, to simplify assumptions needed for later computations. 
Arithmetic done in a positional number system requires keeping track of a carry digit. As such, $k$-digit inputs are unlikely to result in $k$-digit outputs. 
For additon, we first consider the simplest case, a gate with
$m$ single-digit input wires (bounded by $p$), where $m \le p+1$, and two output wires for the digits of the output, which is bounded by $p^2-1$. 
We show in section \ref{AddBasePCosts} that the cost to garble the summation of all $m$ inputs is $2pm - m - 1$. 

Multiplying numbers represented in base-$p$ can be done efficiently by the Karatsuba algorithm, and so we develop the base case of multiplying two
single-digit numbers with a two-digit output in section \ref{MulBasePCosts}, which we found to have a ciphertext cost of $14p-10$. 
We then provide a recurrence relation for the ciphertext cost of a full implementation of the Karatsuba algorithm in section \ref{karatsubaBaseP}.
Finally, we use Karatsuba multiplication to find the cost of multiplying two numbers in a non-native modulus in section \ref{montgomeryCost}.

For squaring, we take a similar divide-and-conquer approach for the initial multiplication done in classical Montgomery multiplication. The Karatsuba
algorithm can be adapted as a recursive call to a unary squaring operation, and as such we provide a recurrence relation for the ciphertext cost of 
Karatsuba squaring in section \ref{karatsubaSquaring}.

\section{The Recursive RNS Method}\label{RNSCosts}
To find the total ciphertext cost of multiplying modulo $N$ in an RNS, we follow the steps outlined in \cite{recursiveRes}.

\subsection{Setup}
We will use English letters to denote public constants and Greek letters to denote private variables; lowercase letters represent variables in the bottom level RNS and uppercase letters represent variables in the top level RNS. We will implement Montgomery multiplication in a setup as follows. 

On the bottom level, we have a left RNS $b = (p_1, p_2, \ldots , p_k)$ with Montgomery constant $m$ equal to the dynamical range $\prod_{i=1}^{k} p_i$, and a right RNS $b' = (p_{k+1}, p_{k+2}, \ldots , p_{2k})$ with dynamical range $m' = \prod_{i=k+1}^{2k} p_i$. 
We define constants 
$c_i = \begin{cases} 
      m/p_i & 1 \leq i \leq k \\
      m'/p_i & k+1 \leq i \leq 2k \\
   \end{cases}$ and take a redundant modulus $p_0 \geq \lceil kt \rceil$ where $t =$ the expansion constant for pseudo-residues. 
   For convenience, we let $b^* = (p_0, p_{k+1}, p_{k+2}, \ldots , p_{2k})$.

On the top level, we have a left RNS $B = (P_1, P_2, \ldots , P_K)$ with Montgomery constant $M$ equal to the dynamical range $\prod_{i=1}^{K} P_i$, and a right RNS $B' = (P_{K+1}, P_{K+2}, \ldots , P_{2K})$ with dynamical range $M' = \prod_{i=K+1}^{2K} P_i$. 
To ensure we can perform arithmetic in every modulus on this level, we require that $0 < (k+2)^2P_i < \min(m, m')$ and $\gcd(P_i, m) = 1\ \forall\ 1 \leq i \leq 2K$. 
We define constants
$C_i = \begin{cases} 
      M/P_i & 1 \leq i \leq K \\
      M'/P_i & K+1 \leq i \leq 2K \\
   \end{cases}$ and take a redundant modulus $P_0 = p_0p_q$ for some $1 \leq q \leq 2k$ with $P_0 \geq \lceil KT \rceil$ where $T =$ the expansion constant for pseudo-residues. 
For convenience, we let $B^* = (P_0, P_{k+1}, P_{k+2}, \ldots , P_{2k})$. 
The modulus $N$ we want to compute in must satisfy $0 < (K+2)^2N < \min(M, M')$ and $\gcd(N, M) = 1$.

\subsection{Bajard-Imbert Montgomery RNS algorithm}

Given $\llbracket \alpha \rrbracket_{b \cup b^*} ,\ \llbracket \beta \rrbracket_{b \cup b^*}$, where $\alpha, \beta \in [0, tn),$ 
we want to compute $$\llbracket \alpha \rrbracket_{b \cup b^*} \otimes_{n,m} \llbracket \beta \rrbracket_{b \cup b^*} = \llbracket \zeta \rrbracket_{b \cup b^*} \text{ where } \zeta \equiv \alpha\beta m^{-1} \mod n \text{ and } \zeta \in [0, tn).$$
When this algorithm is used on its own to perform Montgomery multiplications, $n$ is the non-native modulus. When it is called as a subroutine by the recursive Montgomery RNS algorithm, as we will see in the next section, $n$ is one of the moduli from the second layer.

Before executing the algorithm, we precompute all necessary inverses using the Euclidean algorithm. This does not factor into our cost calculation. 

We begin by computing $$\llbracket \theta \rrbracket_{b \cup b^*} = \llbracket \alpha \rrbracket_{b \cup b^*} \llbracket \beta \rrbracket_{b \cup b^* }$$ by modular multiplication. 
This costs $\sum_{i=0}^{2k} (2p_i - 2)$ ciphertexts if both operands are private variables, and is free if one operand is a public constant. 
For squaring, we can reduce the cost of finding $\llbracket x \rrbracket_{b \cup b^*}^2$
down to $\sum_{i=0}^{2k} p_i - 1$. 

Next, we compute $$[\mu_i]_{p_i} = [\theta]_{p_i} [c_i^{-1}]_{p_i} [-n^{-1}]_{p_i} $$ for each $p_i \in b$ by modular multiplication (which is free because only one operand is a private variable). 
We then project $$[\mu_i]_{p_i} \mapsto \llbracket \mu_i c_i \rrbracket_{b^*}$$ for each $p_i \in b$. 
This costs $(k+1)\sum_{i=1}^k (p_i - 1)$ ciphertexts, since each residue $\mod p_i$ is projected to $k+1$ other moduli. 
We have to move to the right RNS in order to divide by $m$ (the dynamical range of the left RNS), since $m\equiv0\mod p_i \ \forall p_i \in b$, eg. $m$ has no inverse in the left RNS.
Computing $$\llbracket \mu \rrbracket_{b^*} = \dsum_{i=1}^k \llbracket \mu_ic_i \rrbracket_{b^*} \text{ and } \llbracket \zeta \rrbracket_{b^*} = (\llbracket \theta \rrbracket_{b^*} + n \llbracket \mu \rrbracket_{b^*}) \llbracket m^{-1} \rrbracket_{b^*}$$ by modular addition and multiplication is free. 
Now we have to move back to the left RNS, which we will do via generalized base extension using the redundant modulus.
We first compute $$[\eta_i]_{p_i} = [\zeta]_{p_i} [c_i^{-1}]_{p_i}$$ for each $p_i \in b'$ by modular multiplication (which again is free because only one operand is a private variable). 
Next, we project $$[\eta_i]_{p_i} \mapsto [\eta_i]_{p_0}$$ for each $p_i \in b'$ for $\sum_{i=k+1}^{2k} (p_i - 1)$ ciphertexts. 
Computing $$[\omega]_{p_0} = [-m^{-1}]_{p_0}[\zeta]_{p_0} + \dsum_{i=k+1}^{2k}[\gamma_i]_{p_0}[p_i^{-1}]_{p_0}$$ by modular addition and multiplication is free. 
To apply the base extension to the left RNS, we must project $$[\omega]_{p_0} \mapsto \llbracket \omega \rrbracket_b \text{ and } [\eta_i]_{p_i} \mapsto \llbracket \eta_i \rrbracket_b$$ for each $p_i \in b'$, which costs $k(p_0 - 1) + k\sum_{i=k+1}^{2k} (p_i - 1)$ ciphertexts, since the residue $\mod p_0$ and each residue $\mod p_i$ is projected to $k$ other moduli.
We can now complete the base extension by computing $$\llbracket \zeta \rrbracket_b = \dsum_{i=k+1}^{2k} c_i \llbracket \eta_i \rrbracket_b - m' \llbracket \omega \rrbracket_b$$ by modular addition and multiplication for free.
We now have $\llbracket \zeta \rrbracket_{b \cup b^*}$ as desired, for a total cost of $(k+1)\sum_{i=0}^{2k}(p_i -1) - (p_0 - 1),$ plus $\sum_{i=0}^{2k} (2p_i - 2)$ if both operands are private variables.

\subsection{2-layer Montgomery multiplication RNS algorithm}
\subsubsection*{Subroutines}

To implement the 2-layer Montgomery multiplication RNS algorithm, we need several subroutines on the bottom layer. One is the Bajard-Imbert Montgomery multiplication algorithm described above. We will refer to everything in said algorithm after the multiplication $\theta = \alpha\beta$ as Montgomery reduction, denoted $\mathcal{R}_{(n, m)}(\llbracket \theta \rrbracket_{b \cup b^*}) = \llbracket \zeta \rrbracket_{b \cup b^*} $ where $\zeta \equiv \theta m^{-1} \mod n$ and $\zeta \in [0, tn)$ if $\theta \in [0, t^2n^2)$.

We also need the multiply-and-accumulate algorithm, which given $$ \llbracket d_i \rrbracket_{b \cup b^*} ,\ \llbracket \omega_i \rrbracket_{b \cup b^*} \text{ where } d_i \in [0, n) \text{ and } \omega_i \in [0, tn) \text{ for } 1 \leq i \leq s$$ produces $$\mathcal{S}_{n}(d_1, \ldots, d_s; \omega_1, \ldots, \omega_s) = \llbracket \zeta \rrbracket_{b \cup b^*}\text{ where } \zeta \equiv d_1 \omega_1 + \cdots + d_s \omega_s \mod n \text{ and } \zeta \in [0, tn).$$
We can precompute $$\llbracket f_i \rrbracket_{b \cup b^*} = \llbracket\ [m^{\lceil \log_t s \rceil}d_i]_n\ \rrbracket_{b \cup b^*}$$ for $1 \leq i \leq s$, so this does not factor into our total cost. 
We begin by computing $$\llbracket \zeta_i \rrbracket_{b \cup b^*} = \llbracket d_i \rrbracket_{b \cup b^*} \llbracket \omega_i\rrbracket_{b \cup b^*}$$ for $1 \leq i \leq s$.
We will iterate the next part of the algorithm as many times as needed. 

$(*)$ We let $l$ be the number of $\llbracket \zeta_i \rrbracket_{b \cup b^*}$s. If $l > t$, we partition the index set of the $\llbracket \zeta_i \rrbracket_{b \cup b^*}$s into sets $I_1, \ldots, I_r$ so $|I_j| \leq t$ for $1 \leq j \leq r$. Otherwise, we skip to $(**)$.
We then compute $$\llbracket \theta_j \rrbracket_{b \cup b^*} = \dsum_{i \in I_j} \llbracket \zeta_i \rrbracket_{b \cup b^*} \text{ and } \llbracket \zeta_j \rrbracket_{b \cup b^*} = \mathcal{R}_{(n,m)}(\llbracket \theta_j \rrbracket_{b \cup b^*})$$ for $1 \leq j \leq r$. This costs $r \left( (k+1)\sum_{i=0}^{2k}(p_i -1) - (p_0 - 1) \right)$ ciphertexts. We return to $(*)$ and repeat. 

$(**)$ We can now compute $$\llbracket \theta \rrbracket_{b \cup b^*} = \dsum_{j=1}^l \llbracket \zeta_j \rrbracket_{b \cup b^*} \text{ and } \llbracket \zeta \rrbracket_{b \cup b^*} = \mathcal{R}_{(n,m)}(\llbracket \theta \rrbracket_{b \cup b^*})$$
for $(k+1)\sum_{i=0}^{2k}(p_i -1) - (p_0 - 1)$ ciphertexts. This gives us a total cost of \\ $\left( \sum_{i=1}^{\lfloor \log_t s \rfloor} \frac{s}{t^i} + 1 \right) \left( (k+1)\sum_{i=0}^{2k}(p_i -1) - (p_0 - 1) \right)$ ciphertexts. \\

\subsubsection*{Implementation}

With these three subroutines, we can implement Montgomery multiplication in a 2-layer RNS. For convenience, we write $(\alpha)_{P_0}$ to denote $([\alpha]_{p_0}, [\alpha]_{p_q})$ and $(\alpha)_{P_i}$ to denote $\llbracket \alpha \mod P_i \rrbracket_{b \cup b^*}$ for $P_i \in B \cup B'$.
Given $$( (\alpha)_{P_0}, (\alpha)_{P_1}, \ldots, (\alpha)_{P_{2K}}), ( (\beta)_{P_0}, (\beta)_{P_1}, \ldots, (\beta)_{P_{2K}}) \text{ where } \alpha, \beta  \in [0, TN),$$ we want to compute $$( (\zeta)_{P_0}, (\zeta)_{P_1}, \ldots, (\zeta)_{P_{2K}}) \text{ where } \zeta \equiv \beta\gamma M^{-1} \mod N \text{ and }\zeta \in [0, TN).$$
We begin by computing $$(\theta)_{P_i} =(\beta)_{P_i} \otimes_{P_i, m} (\gamma)_{P_i} $$ for each $P_i \in B \cup B'$. This costs $2K \left( \sum_{i=0}^{2k} (2p_i - 2) + (k+1)\sum_{i=0}^{2k}(p_i -1) - (p_0 - 1) \right) $ ciphertexts, because we have to do $2K$ Montgomery multiplications on the bottom level. These multiplications must use the Montgomery algorithm to ensure that the results are pseudo-residues $\mod P_i$, since regular multiplication wouldn't guarantee any bounds in terms of $P_i$.
The redundant modulus $P_0$ is the product of two moduli from the lower level, so we have to handle arithmetic $\mod P_0$ differently. We can compute $$(\theta)_{P_0} = (\alpha)_{P_0} (\gamma)_{P_0}$$ by modular multiplication, for $2(p_0 + p_q) - 4$ ciphertexts.
Having precomputed $(A_i)_{P_i} = \llbracket\ [ -N^{-1} C_i^{-1} m^2 ]_{P_i}\ \rrbracket_{b \cup b^*}$ for $P_i \in B$, we can now compute $$(\mu_i)_{P_i} = (\theta)_{P_i} \otimes_{P_i, m} (A_i)_{P_i}$$ for each $P_i \in B$, for $K \left( (k+1)\sum_{i=0}^{2k}(p_i -1) - (p_0 - 1) \right)$ ciphertexts.
We then compute $$(\zeta)_{P_0} = (\theta)_{P_0} (M^{-1})_{P_0} + \dsum_{i=1}^K (\mu_i)_{P_0}(NP_i^{-1})_{P_0}$$ for free by modular addition and multiplication.
With precomputations of $(D_{i,0})_{P_i}  = \llbracket\ [M^{-1}m]_{P_i}\ \rrbracket_{b \cup b^*}$ for $P_i \in B'$ and $(D_{i,j})_{P_i} = \llbracket\ [N P_j^{-1} ]_{P_i}\ \rrbracket_{b \cup b^*}$ for $P_i \in B', P_j \in B$, we use the multiply-and-accumulate subroutine to compute $$(\zeta)_{P_i} = \mathcal{S}_{P_i}((D_{i,0})_{P_i}, (D_{i,1})_{P_i}, \ldots, (D_{i,K})_{P_i}; (\theta)_{P_i}, (\mu_1)_{P_1}, \ldots, (\mu_K)_{P_K})$$ for each $P_i \in B'$. 
This costs $K \left( \sum_{i=1}^{\lfloor \log_t (K+1) \rfloor} \frac{K+1}{t^i} + 1 \right) \left( (k+1)\sum_{i=0}^{2k}(p_i -1) - (p_0 - 1) \right)$ ciphertexts, since we have $K+1$ inputs to the subroutine and run it $K$ times.
We now have our desired result in the upper right RNS and the redundant modulus, but need to use base extension to move it back to the upper left RNS.
Given precomputed $(E_i)_{P_i} = \llbracket\ [ C_i^{-1} m ]_{P_i}\ \rrbracket_{b \cup b^*}$ for $P_i \in B$, we then compute $$(\alpha_i)_{P_i} = (\zeta)_{P_i} \otimes_{P_i, m} (E_i)_{P_i}$$ for each $P_i \in B'$. 
This costs $K \left( (k+1)\sum_{i=0}^{2k}(p_i -1) - (p_0 - 1) \right)$ ciphertexts, since we are doing $K$ Montgomery multiplications on the bottom layer (for the same reasoning as stated above).
We now compute $$(\omega)_{P_0} = (\zeta)_{P_0}(-M'^{-1})_{P_0} + \dsum_{i=K+1}^{2K} (\alpha_i)_{P_0}(P_i^{-1})_{P_0}$$ by modular addition and multiplication for free. We don't need to project the $\alpha_i$s to get each $(\alpha_i)_{P_0}$ because $P_0$ is the product of moduli on the bottom layer, and each $\alpha_i$ is already represented as residues in every bottom layer moduli. However, to go in the other direction we do need to project $$(\omega)_{P_0} \mapsto \llbracket \omega \rrbracket_{b \cup b^*}$$ for $(2k+1)(P_0 - 1)$ ciphertexts, since we are projecting to $2k+1$ moduli. 
To complete the base extension to the upper left RNS, we need precomputations of $(D_{i,0})_{P_i}  = \llbracket\ [-M']_{P_i}\ \rrbracket_{b \cup b^*}$ for $P_i \in B$ and $(D_{i,j})_{P_i} = \llbracket\ [C_j]_{P_i}\ \rrbracket_{b \cup b^*}$ for $P_i \in B, P_j \in B'$ plugged into the multiply-and-accumulate subroutine to compute $$(\zeta)_{P_i} = \mathcal{S}_{P_i}((D_{i,0})_{P_i}, (D_{i,K+1})_{P_i}, \ldots, (D_{i,2K})_{P_i}; \llbracket \omega \rrbracket_{b \cup b^*}, (\alpha_{K+1})_{P_{K+1}}, \ldots, (\alpha_{2K})_{P_{2K}})$$ for each $P_i \in B$.
This costs $K \left( \sum_{i=1}^{\lfloor \log_t (K+1) \rfloor} \frac{K+1}{t^i} + 1 \right) \left( (k+1)\sum_{i=0}^{2k}(p_i -1) - (p_0 - 1) \right)$ ciphertexts, since again we have $K+1$ inputs to the subroutine and run it $K$ times.
We now have $( (\zeta)_{P_0}, (\zeta)_{P_1}, \ldots, (\zeta)_{P_{2K}})$ as desired, for a total cost of
\begin{multline*}
2(p_0 + p_q) - 4 + 2K\sum_{i=0}^{2k}(2p_i - 2) + (2k+1)(P_0 - 1) \\ + K \left( 3+2 \left( \sum_{i=1}^{\lceil \log_t(K+1) \rceil} \frac{K+1}{t^i} + 1 \right) \right) \left( (k+1)\sum_{i=0}^{2k}(p_i-1) - (p_0-1) \right)
\end{multline*} ciphertexts.

\section{Circuits With Base-$p$ Input Wires}

\subsection{Basic addition and multiplication}
\subsubsection{Cost of addition}\label{AddBasePCosts}
Adding $m$ residues in base $p$ (where $m \leq p + 1$ to ensure the overflow is limited to a single digit) costs $2mp - m - 1$ ciphertexts. First, the residues are added $\mod p$ for free to find the less significant digit of the sum. Then each residue is cast to $\mathbb{Z}_{mp}$ for $p-1$ ciphertexts each (for a total of $m(p-1)$ ciphertexts), and the projected residues are added $\mod mp$ for free. Lastly, the sum mod $mp$ is projected to the appropriate carry digit in $\mathbb{Z}_p$ for $mp-1$ ciphertexts.
\subsubsection{Cost of single-digit multiplication}\label{MulBasePCosts}
In \cite{BMR16}, a discrete-logarithm technique for multiplying in $\Z_p$ (for prime $p$) is described. We define a discrete logarithm
$d: \Z_p\setminus \{0\} \to \Z_{p-1}$, 
where $d: g^a \mapsto a$ for some fixed multiplicative generator $g\in\Z_p\setminus \{0\}$. Then, since $g^ag^b = g^{a+b}$, 
we can multiply two private inputs $x, y \in\Z_p\setminus \{0\}$ by
adding $d(x) + d(y)$ without any ciphertext cost, and taking $g^{d(x) + d(y)} = xy$.

Suppose now that we wish to multiply two private variable residues $[A]_p$ and $[B]_p$. 
We can compute the less significant digit of the product by multiplying in $\mathbb{Z}_p$ for $2p-2$ ciphertexts, but computing the carry digit is more involved. First we check if either residue is $0$ for $2p$ ciphertexts. If not, we have $$[A]_p = g^a - \alpha p,\ [B]_p = g^b - \beta p$$ (equalities in $\mathbb{Z}$) for $g$ a generator in $\mathbb{Z}_p$. We can project $[A]_p$ and $[B]_p$ to $a, \alpha, b, \beta$ for $4p-4$ ciphertexts. Then $$[A]_p \cdot [B]_p = (g^a - \alpha p)(g^b - \beta p) = g^{a+b} + p \left( \alpha\beta p - g^b \alpha - g^a \beta \right).$$ If we compute $c = a+b$ in $\mathbb{Z}_{2p-1}$ for free, we have $$[A]_p \cdot [B]_p = [g^c]_p + p \left(\left \lfloor \frac{g^c}{p} \right \rfloor + \alpha\beta p - g^b \alpha - g^a \beta \right).$$ We can project $c$ to $\left \lfloor \frac{g^c}{p} \right \rfloor$ for $2p-2$ ciphertexts. Since the carry digit is bounded by $p$ (because the product is bounded by $p^2$), we can compute it in $\mathbb{Z}_p$ as $$\left \lfloor \frac{g^c}{p} \right \rfloor + \alpha\beta p - g^b \alpha - g^a \beta \equiv \left \lfloor \frac{g^c}{p} \right \rfloor - [B]_p \alpha - [A]_p \beta \mod p.$$ 
This requires two $\mathbb{Z}_p$ multiplications for $2p-2$ ciphertexts each, bringing the total to $14p-10$ ciphertexts to multiply single digit numbers in base-$p$.

Multiplication by a public constant is significantly cheaper, since all multiplication in $\mathbb{Z}_p$ is free and we have fewer lookups to do. If we let $[B]_p$ be a public constant, we only need $p-1$ ciphertexts to check if $[A]_p =0$, $2p-2$ ciphertexts to lookup $a$ and $\alpha$, and $2p-2$ ciphertexts to lookup $\left \lfloor \frac{g^c}{p} \right \rfloor$, for a total of $5p-5$ ciphertexts.

\subsection{Implementing Karatsuba multiplication algorithm}\label{karatsubaBaseP}
For multiplication of many-digit base $p$ numbers, we resort to the Karatsuba algorithm. Since our implementation of Karatsuba requires fast negation, we use ``$p$'s complement'' notation. An $m$-digit base $p$ number is represented by $m$ $\mod p$ wires and one $\mod 2$ wire which keeps track of the sign. Like in two's complement, negation requires flipping every digit and adding $1$. To flip a digit on a $\mod p$ wire, we send $a \mapsto p - 1 - a$. Flipping digits on mod $2$ wires is just a standard bit flip satisfying. Both procedures are free.

Suppose we wish to multiply two $m$-digit base $p$ numbers $X$ and $Y$. We can split each number in half so we have $$X = X_1 p^{\lceil m/2 \rceil} + X_0,\ Y = Y_1 p^{\lceil m/2 \rceil}  + Y_0,$$ where $X_1, Y_1$ are $\lceil m/2 \rceil$-digit base $p$ numbers and $X_0, Y_0$ are $\lfloor m/2 \rfloor$-digit base $p$ numbers. Karatsuba's algorithm allows us to perform only $3$ (instead of $4$ as required by the schoolbook algorithm) multiplications of these parts. Each multiplication is recursively performed with the Karatsuba algorithm, until the operands have few enough digits that it is more efficient to use schoolbook multiplication (multiplying every digit of one operand by every digit of the other and adding all the products). The first two multiplications are $$A = X_0 \cdot Y_0,\ C = X_1 \cdot Y_1,$$ of $\lceil m/2 \rceil$ and $\lfloor m/2 \rfloor$ digits, respectively. The next multiplication is more involved. First we compute $$X' = X_0 - X_1,\ Y' = Y_1 - Y_0$$ for $12 \lceil m/2 \rceil p - 8\lceil m/2 \rceil$ ciphertexts. Next we take the absolute value of $X'$ and $Y'$ (by negating if the sign bit is $1$) for $8 \lceil m/2 \rceil p - 6\lceil m/2 \rceil$ ciphertexts. Then we can perform our last multiplication $$B = |X'| \cdot |Y'|,$$ of $\lceil m/2 \rceil$ digits. If $X'$ and $Y'$ originally had different signs, we now negate $B$ for $4 \lceil m/2 \rceil p - 3\lceil m/2 \rceil$ ciphertexts. Lastly, we compute our desired product $$X \cdot Y = C p^m + C p^{\lceil m/2 \rceil} + B p^{\lceil m/2 \rceil} + A p^{\lceil m/2 \rceil} + A$$ for $10mp - 6m + 4 \lceil m/2 \rceil p - 3\lceil m/2 \rceil$ ciphertexts. Thus we have a recurrence $$T(m) = 2T(\lceil m/2 \rceil) + T(\lfloor m/2 \rfloor) + 10mp - 6m + 28 \lceil m/2 \rceil p - 20\lceil m/2 \rceil$$ which, by the Master Theorem, gives us an asymptotic bound of $T(m) = \Theta(m^{\log_2 3}).$

Many-digit multiplication by a public constant uses the exact same Karatsuba algorithm, but is cheaper because the base case of the recursion is a schoolbook multiplication by a constant, which uses single-digit multiplication by a constant ($5p-5$ ciphertexts) instead of single-digit multiplication of private variables ($14p-12$ ciphertexts).

\subsection{Implementing classical Montgomery arithmetic}\label{montgomeryCost}
Now that we have all the necessary arithmetic operations, we can implement classical Montgomery multiplication in base $p$. We precompute $$(q)_p = (N^{-1} \mod p^k)_p.$$ Multiplying $$(\theta)_p = (\alpha)_p(\beta)_p$$ requires a Karatsuba multiplication of $k$ digits. Although $(\theta)_p$ has $2k$ digits, since the next multiplication $$(\mu)_p = -(\theta)_p(q)_p \mod p^k$$ is over $\mathbb{Z}_{p^k}$, we can ignore the $k$ more significant digits of $(\theta)_p$, and so this requires only a Karatsuba multiplication by a constant of $k$ digits. The next multiplication $$(\gamma)_p = (\mu)_p(N)_p$$ is also a Karatsuba multiplication by a constant of $k$ digits. Computing $$(\zeta)_p = ( (\theta)_p + (\gamma)_p) \gg k$$ requires a $12kp - 2p - 8k + 1$ addition and a free bit shift. Lastly, we compute $$(\zeta')_p = (\zeta)_p - (N)_p$$ for $6kp - 4k$ ciphertexts. If the sign bit of $(\zeta')_p$ is $0$ (which means $(\zeta)_p \geq (N)_p$), we set $(\zeta)_p = (\zeta')_p$.

\subsection{Squaring in Base-$p$}\label{karatsubaSquaring}
For the special case of squaring a single $m$-digit base $p$ number $X$, we take a slightly different approach. We still split $X$ in half and compute $$A = X_0 \cdot X_0,\ C = X_1 \cdot X_1,$$ two recursive squarings of $\lfloor m/2 \rfloor$ and $\lceil m/2 \rceil$ digits, respectively. For the third multiplication, we compute $$X' = X_1 + X_0,\ B = X' \cdot X'$$ for $6 \lceil m/2 \rceil p - 2p - 4\lceil m/2 \rceil + 1$ ciphertexts, plus the cost of squaring a $\lceil m/2 \rceil + 1$ digit number. Lastly, we compute our desired square $$X^2 = C p^m - C p^{\lceil m/2 \rceil} + B p^{\lceil m/2 \rceil} - A p^{\lceil m/2 \rceil} + A$$ for $10mp - 6m + 4 \lceil m/2 \rceil p - 3\lceil m/2 \rceil + 6p - 4$ ciphertexts. Thus we have a recurrence $$S(m) =  S(\lfloor m/2 \rfloor) + S(\lceil m/2 \rceil) + S(\lceil m/2 \rceil + 1) + 10mp - 6m + 10 \lceil m/2 \rceil p - 7\lceil m/2 \rceil +4p - 3.$$\\

\section{Performance}

\subsection{Montgomery multiplication}

\begin{center}
Performance of Bajard-Imbert RNS algorithm \\ versus our base $p$ algorithm for Montgomery multiplication: \\
\end{center}

{
\centering
\begin{tabular}{r || l | l | l | l | l | l}
N bit len   & RNS         & Base 2      & Base 3      & Base 5      & Base 7      & Base 11    \\
\hline 
\rule{0pt}{2.6ex}
139 & $ 1.48*10^5 $ & $ 3.35*10^5 $ & $ 3.90*10^5 $ & $ 3.77*10^5 $ & $ 4.15*10^5 $ & $ 4.90*10^5 $ \\
350 & $ 1.30*10^6 $ & $ 1.40*10^6 $ & $ 1.68*10^6 $ & $ 1.71*10^6 $ & $ 1.80*10^6 $ & $ 2.18*10^6 $ \\
585 & $ 4.68*10^6 $ & $ 3.35*10^6 $ & $ 3.92*10^6 $ & $ 3.78*10^6 $ & $ 4.23*10^6 $ & $ 5.03*10^6 $ \\
835 & $ 1.16*10^7 $ & $ 5.50*10^6 $ & $ 7.14*10^6 $ & $ 6.91*10^6 $ & $ 7.56*10^6 $ & $ 8.70*10^6 $ \\
1364 & $ 4.17*10^7 $ & $ 1.26*10^7 $ & $ 1.51*10^7 $ & $ 1.54*10^7 $ & $ 1.62*10^7 $ & $ 1.92*10^7 $ \\
1924 & $ 1.04*10^8 $ & $ 2.21*10^7 $ & $ 2.64*10^7 $ & $ 2.61*10^7 $ & $ 2.88*10^7 $ & $ 3.43*10^7 $ \\
2504 & $ 2.09*10^8 $ & $ 3.27*10^7 $ & $ 4.05*10^7 $ & $ 4.10*10^7 $ & $ 4.14*10^7 $ & $ 5.05*10^7 $ \\
\end{tabular} \\
}
We found that the base $p$ arithmetic-based Montgomery multiplication implementation outperformed the Bajard-Imbert RNS algorithm for all values of $N$ with bit length $ \gtrsim 500$. This was surprising to us because the RNS algorithm took full advantage of free addition and we expected it to perform better than an implementation in a positional number system. However, the overhead associated with converted back and forth between the left and right RNS, including base extension, meant that the RNS algorithm was less efficient overall.\vspace*{-20pt}

\subsection{Montgomery squaring}
\begin{center}

Performance of Bajard-Imbert RNS algorithm \\ versus our base $p$ algorithm for Montgomery squaring: \\
\end{center}

{
\centering
\begin{tabular}{r || l | l | l | l | l | l}
N bit len  & RNS         & Base 2      & Base 3      & Base 5      & Base 7      & Base 11    \\
\hline
\rule{0pt}{2.6ex}
139 & $ 1.43*10^5 $ & $ 3.40*10^5 $ & $ 3.30*10^5 $ & $ 3.14*10^5 $ & $ 3.37*10^5 $ & $ 4.03*10^5 $\\
350 & $ 1.28*10^6 $ & $ 1.45*10^6 $ & $ 1.43*10^6 $ & $ 1.44*10^6 $ & $ 1.51*10^6 $ & $ 1.78*10^6 $ \\
585 & $ 4.62*10^6 $ & $ 3.42*10^6 $ & $ 3.32*10^6 $ & $ 3.20*10^6 $ & $ 3.48*10^6 $ & $ 4.15*10^6 $ \\
835 & $ 1.15*10^7 $ & $ 5.72*10^6 $ & $ 6.08*10^6 $ & $ 5.72*10^6 $ & $ 6.31*10^6 $ & $ 7.25*10^6 $ \\
1364 & $ 4.15*10^7 $ & $ 1.29*10^7 $ & $ 1.29*10^7 $ & $ 1.29*10^7 $ & $ 1.36*10^7 $ & $ 1.58*10^7 $ \\
1924 & $ 1.03*10^8 $ & $ 2.24*10^7 $ & $ 2.27*10^7 $ & $ 2.16*10^7 $ & $ 2.38*10^7 $ & $ 2.84*10^7 $ \\
2504 & $ 2.09*10^8 $ & $ 3.38*10^7 $ & $ 3.42*10^7 $ & $ 3.42*10^7 $ & $ 3.45*10^7 $ & $ 4.17*10^7 $ \\
\end{tabular} \\
}
We found that in the special case of squaring, the Bajard-Imbert RNS algorithm experienced only a slight improvement in ciphertext cost, likely because the inputs being the same did not reduce the overhead needed for Montgomery reduction at all, while the base $p$ implementations saw a more significant improvement. We also found that base $3$ and base $5$ performed better than binary for some values of N. 
\bigskip
\bigskip
\bibliographystyle{plain}

\break
\bibliography{mybib}

\end{document}